\documentclass{article}
\usepackage{textcomp}
\usepackage{color}
\usepackage{pdfpages}
\usepackage{url}
\usepackage[a4paper, margin=2in]{geometry}
 \DeclareUnicodeCharacter{03A9}{$\Omega$}
 \DeclareUnicodeCharacter{03B2}{$\beta$}

\begin{document}

\title{Application of dual-tree complex wavelet transform for spectra background reduction}
\date{}

\maketitle

\author{Kazimierz Skrobas*}, 
\author{Kamila Stefańska-Skrobas}, 
\author{Cyprian Mieszczyński}, 
\author{Renata Ratajczak}\\

National Centre for Nuclear Research, A. Soltana 5, 05-400 Otwock-Swierk, Poland \\

\


\textbf{*coresponding author:} Kazimierz.Skrobas@ncbj.gov.pl	\\

\begin{abstract}

		This paper presents a method for unique background removal in experimental data processing using the Dual-Tree Complex Wavelet Transform (DTCWT) algorithm. The proposed technique is based on discrete wavelet theory (DWT) and overcomes numerous obstacles encountered in information retrieval that are present in commonly used numerical techniques, particularly fitting or filtering methods. The DTCWT also outperforms methods based on the Fourier Transform. The method is universal, allowing the analysis of an arbitrarily selected data range and its position in time. Furthermore, it meets the specific requirements of signal analysis and optimization, including signal preservation and bias reduction. This paper discusses the implementation of an algorithm for background reduction, enabling the extraction and enhancement of valuable information from spectra. The capabilities of wavelets for spectral data processing are demonstrated using two significantly different types of spectra: X-ray powder diffraction and photoluminescence measured for the $Ga_{2}O_{3}$ crystal. Issues typical of DWT applications that are important for robust and reliable data processing—such as the choice of wavelet family and the number of decomposition levels—are also discussed. The method is available as a software package for background reduction.

\end{abstract}

\section{Motivations}
Experimental data always consists of informative and non-informative parts. The last one is a compilation of different types of distortions and often makes extracting the valuable part a challenging job. Typically, the Fourier Transform (FT) is employed for distortion separation  \cite{Wahab, Peterson, Neves, Cole}. The FT decomposes the signal in the frequency domain into slow, medium, and fast-changing components, where: (i) the lowest frequencies are responsible for background, (ii) the medium band is usually an interesting region for further analysis, and  (iii) the highest frequencies' region originates from bursts of short-duration impulses (high-frequency noise). By suppressing the extreme frequencies and leaving the medium sub-band untouched, it is possible to synthesize a signal free of distortion.
	
	The disadvantage of the FT techniques is the sensitivity to the range of data taken for analysis, which means that regions that are too short generate spurious ripples near the correct peaks in the frequency domain. Another drawback is overlapping frequency ranges and aliasing artifacts. All of the above are sources of errors in FT analysis of signals consisting of components with frequencies dependent on time, e.g., human speech, trends, chirp-like sounds, or any non-stationary deviations.
	
	Wavelet-based techniques can address the above issues. They originate from the theory of orthogonal function systems \cite{haar}, which has been transformed to the Wavelet Theory (WT) by \cite{gabor,morlet,daub0,daub1}. Contrary to FT, wavelets are both well localized in time and frequency domains, which allows for a significantly better introduction of algorithms in signal processing. The high effectiveness of the WT has led to its acceptance in numerous fields. It outperforms commonly used techniques for signal processing, like the Windowed Fourier Transform, and other domains for image processing or data compression.

        Wavelet-based approaches have recently been used as techniques for background removal. This is particularly important when the level of slowly varying components is comparable to, or even higher than, the level of useful information contained in the signal. For example, this issue was addressed by \cite{hupfel} et al., who used a WT-based algorithm for the decontamination of fluorescence microscopy images from both low- and high-frequency noise. Furthermore, they demonstrated that this approach helps mitigate hardware-related issues that may significantly affect the final results.

The first background removal algorithm was proposed by Galloway et al.~\cite{galloway} and was applied to data processing in surface-enhanced Raman spectroscopy (SERS). The Galloway algorithm was later further developed by Zhao \cite{fzhao} and Cotret \cite{rene0} through the introduction of the Dual-Tree Complex Wavelet Transform (DTCWT) \cite{dtcw}, which originated from efforts to improve the Discrete Wavelet Transform (DWT). The DTCWT enabled a significant reduction of oscillations in the vicinity of sharp peaks and the elimination of aliasing artifacts. The use of DTCWT for background subtraction from energy-dispersive X-ray fluorescence (EDXRF) spectra was demonstrated in \cite{fzhao}, for ultrafast electron scattering (UES) 1D spectra in \cite{rene0}, and for 2D scattering patterns in \cite{rene1}. In the latter case, the removal of image hotspots using the Discrete Wavelet Transform was also demonstrated.

\section{Wavelet Transformations}

Similarly to the Fourier Transform, two techniques can be distinguished, corresponding to signals that are continuous or discrete in time. The Continuous Wavelet Transform (CWT) of a $f(t)$ function given by \cite{morlet} is the following:
\begin{equation}
	\label{equ:cwt}
	Wf(b,a)=\int_{-\infty}^{+\infty} f(t)\psi_{ab}^{\star}(t)dt
\end{equation}
where $\psi_{ab}(t)$ is a wavelet function (WF):
\begin{equation}
		\label{equ:cdwt}
	\psi_{ab}(t)=\frac{1}{\sqrt a}\psi\left(\frac{t-b}{a} \right ) \quad a \in \mathbf{R}^+, b \in \mathbf{R}
\end{equation}
and the Inverse Continuous Wavelet Transform (ICWT):
\begin{equation}
	\label{equ:icwt}
	f(t)=\frac{1}{C} \int_{0}^{+\infty}\int_{-\infty}^{+\infty} Wf(b,a)\frac{1}{a}\psi_{ab}(t) dadb
\end{equation}

The  \textit{a}  controls WF dilation and is called a \textit{scale parameter}; the \textit{b} is called a shift parameter, giving the wavelet location in time; C is a wavelet-dependent constant. In general, WF are complex functions, non-orthogonal to themselves, but orthogonal to the dual basis. The   Equ.\ref{equ:cwt} can be interpreted as a convolution of the input $f(t)$ function with a WF.

The CWT produces a very detailed image of a signal with a significant amount of redundant information. For numerous purposes, such as multilevel decomposition, it is feasible to substitute CWT with a more concise view. In this case, orthogonal, real-value functions are selected, and the continuous \textit{a,b} parameters are replaced by values from a discrete set. Typically, in such a dyadic decomposition, the scale and shift parameters are powers of 2. This approach, called a Discrete Wavelet Transform (DWT), transforms  a discrete signal $f[n]$ as follows:
\begin{equation}
	\label{equ:dwt}
DWT(j,k)=\sum_{n}f[n]\psi_{jk}[n]
\end{equation}
\begin{equation}
	\label{equ:wdwt}
	\psi_{jk}[n]=\frac{1}{\sqrt {2^j}}\psi\left(\frac{k-n2^j}{2^j} \right)  
\end{equation}
and the Inverse Discrete Wavelet Transform (IDWT):
\begin{equation}
	f[n]=\sum_{jk}DWT(j,k)\psi_{jk}[n]
\end{equation}

for integers $j,k,n \in \mathbf{Z}$. Here, the ${j}$ parameter controls the scale and selects the subband of frequencies covered by a wavelet.  The lowest range of frequencies is represented by the highest $j$ values. The ${j}$ parameter is the subject of adjusting to control the number of decomposition levels. The maximum value of levels is given by $L_{max}=log_2{N}$, where N stands for the number of data points. 

\begin{figure}
	\centering
    \includegraphics[width=0.5\linewidth]{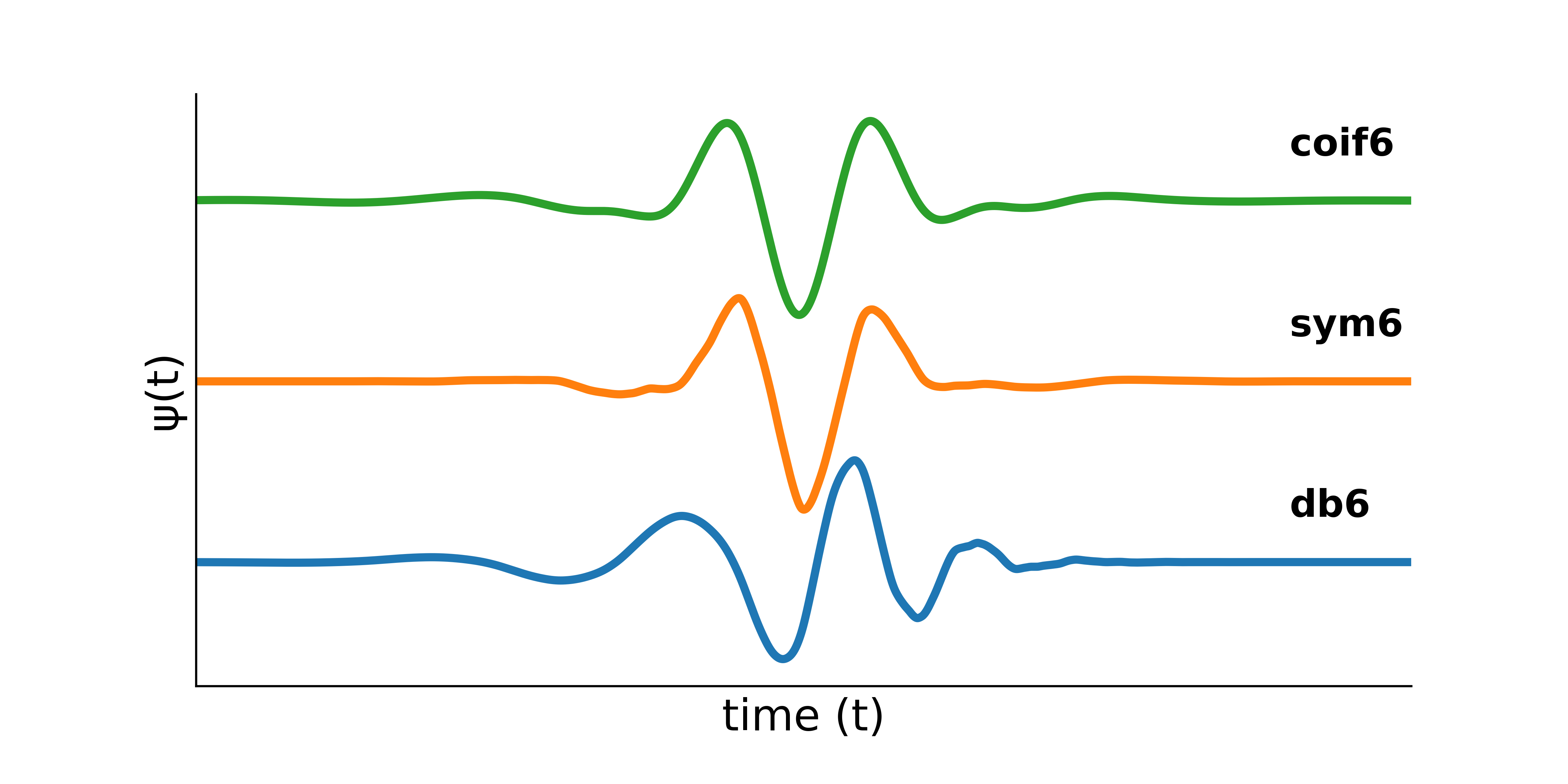}
	\caption{Example wavelets used in DWT selected from 3 families, i.e. Daubechies (db6), Symlet (sym6), Coiflet (coif6) of 8th level (the number followed by the name of the wavelet stands for the corresponding filter's order) }
	\label{fig:wavExample}
\end{figure}

There are numerous collections, called families, from which one can select \textit{mother} wavelet (MW), which is next a subject of modification by adjusting the \textit{i,j} parameters.  For example, the orthogonal $\psi(n)$ functions for DWT are shown in Fig.\ref{fig:wavExample} from Daubechies \textit{db}, Symlet \textit{sym}, and Coiflet \textit{coif} families.  The \textit{db} family first was given by Daubechies \cite{daub0}; the Symlet and Coiflet wavelets are modified Daubechies wavelets as to obtain greater symmetry and number of vanishing moments (smoothness) \cite{daub2} of an MW.

\begin{figure}[ht]
	\centering
	\includegraphics[width=0.35\linewidth]{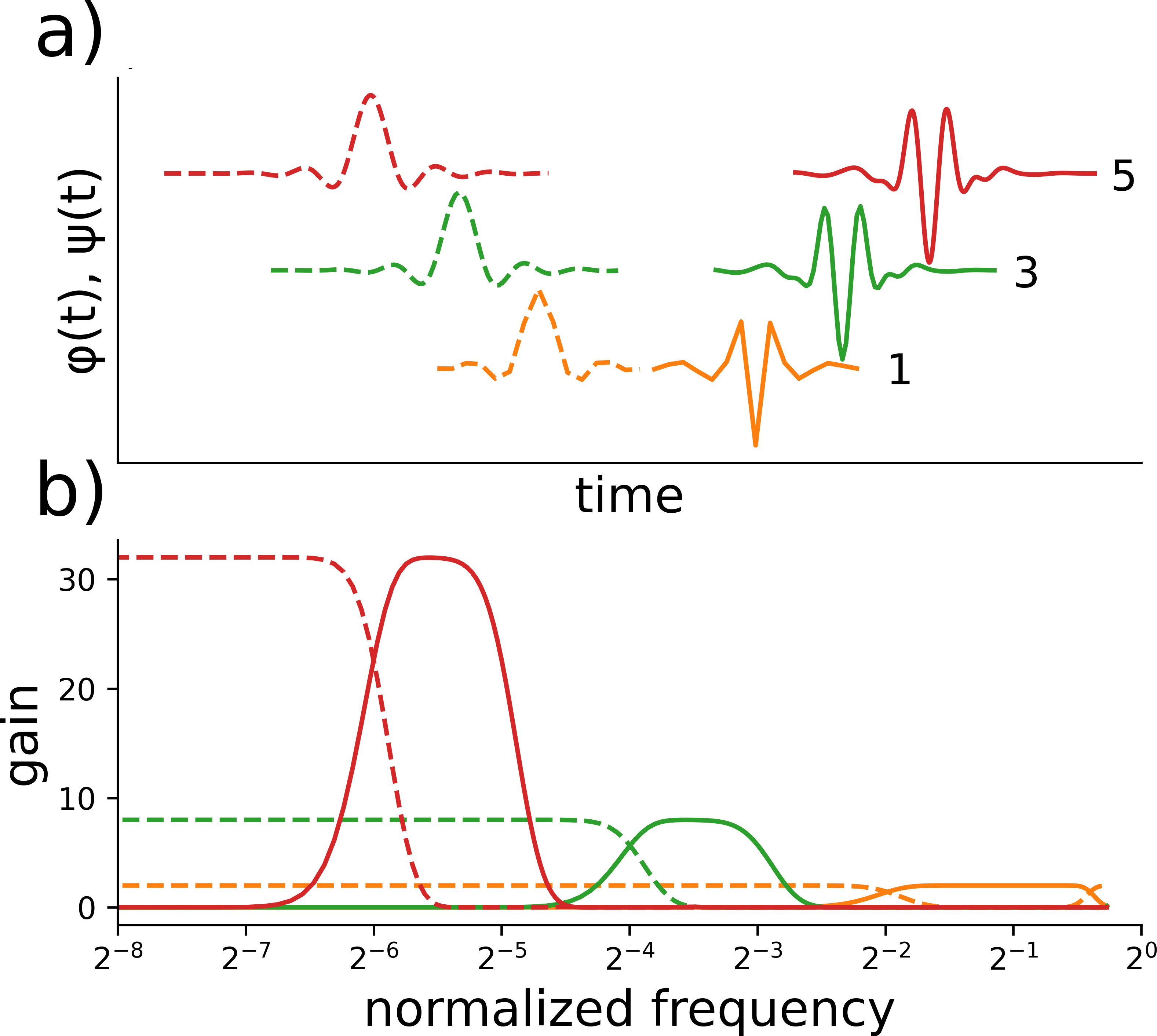}
	\caption{a) The scaling (dashed lines) and wavelet  (solid lines)  functions from the Coiflet family (coif8); the number stands for a decomposition level. b) Corresponding frequency characteristics}
	\label{fig:wavfreqch}
\end{figure}

The common feature of all wavelets, in contrast to trigonometric functions used in the Fourier transform, is that they are finite and well localized both in time, see Fig.\ref{fig:wavfreqch}a) and frequency, Fig.\ref{fig:wavfreqch}b) (solid lines). They can be treated as a band-pass filter, with a central frequency proportional to $2^{-j}$ and gain and bandwidth proportional to $2^j$. The lowest frequencies range with a constant component are supplemented by \textit{scaling function} (SF), see in see Fig.\ref{fig:wavfreqch}a,b) (dashed line),  which acts as a low-pass filter. The collection of WF and the corresponding SF functions is called a bank.

The presented method of a signal decomposition based on filter banks is called multiresolution analysis (MA) and was proposed by Mallat\cite{mallat0}. The original MA uses real wavelets only and has been extended to complex wavelets, what become a foundation of the DTCWT \cite{dtcw} algorithm. The DTCWT addresses issues such as oscillations, shift variance, and aliasing.
It also found an application in the background removal software and in the results presented here.

\section{Results}

$\beta-Ga_{2}O_{3}$ is one of the most promising wide-bandgap semiconductors, meeting the demands of modern applications in high-power electronics, optoelectronics, and solar-blind detectors \cite{ga2o3}. Moreover, it is a radiation-resistant material, meaning that devices made from $\beta-Ga_{2}O_{3}$ can operate in radiation-intensive environments, such as space. As a result, studying radiation-induced defects and other phenomena in this material is crucial. In addition, $\beta-Ga_{2}O_{3}$ is considered to be an excellent host material for rare earth (RE) ions, which can provide their more efficient light emission compared to other wide-bandgap semiconductors. Hence, such systems are intensively studied \cite{ga2o3,MS1,MS2}. However, the structural and optical response of the RE dopant is very subtle due to its low concentration in the matrix. Therefore, it is essential to eliminate any electronic disturbances contributing to the experimental spectra or extract the information coming from the very weak signals of the RE on a strong background.
In the present work, the WT-based algorithms have been used for background analyses of X-ray and photoluminescence (PL) spectra processing obtained for $\beta-Ga_{2}O_{3}$ crystals implanted at room temperature with $150\ keV$ rare earth ions like Yb, and Eu with fluency $1\times10^{15}\ 1/{cm^{2}}$.

The obtained spectra significantly differ in terms of peak properties (width, height, relative distances), background levels, and high-frequency noise levels, which allowed testing the application of the DTCWT algorithm on different conditions.

\begin{figure}
    \centering
	\includegraphics[width=0.5\linewidth]{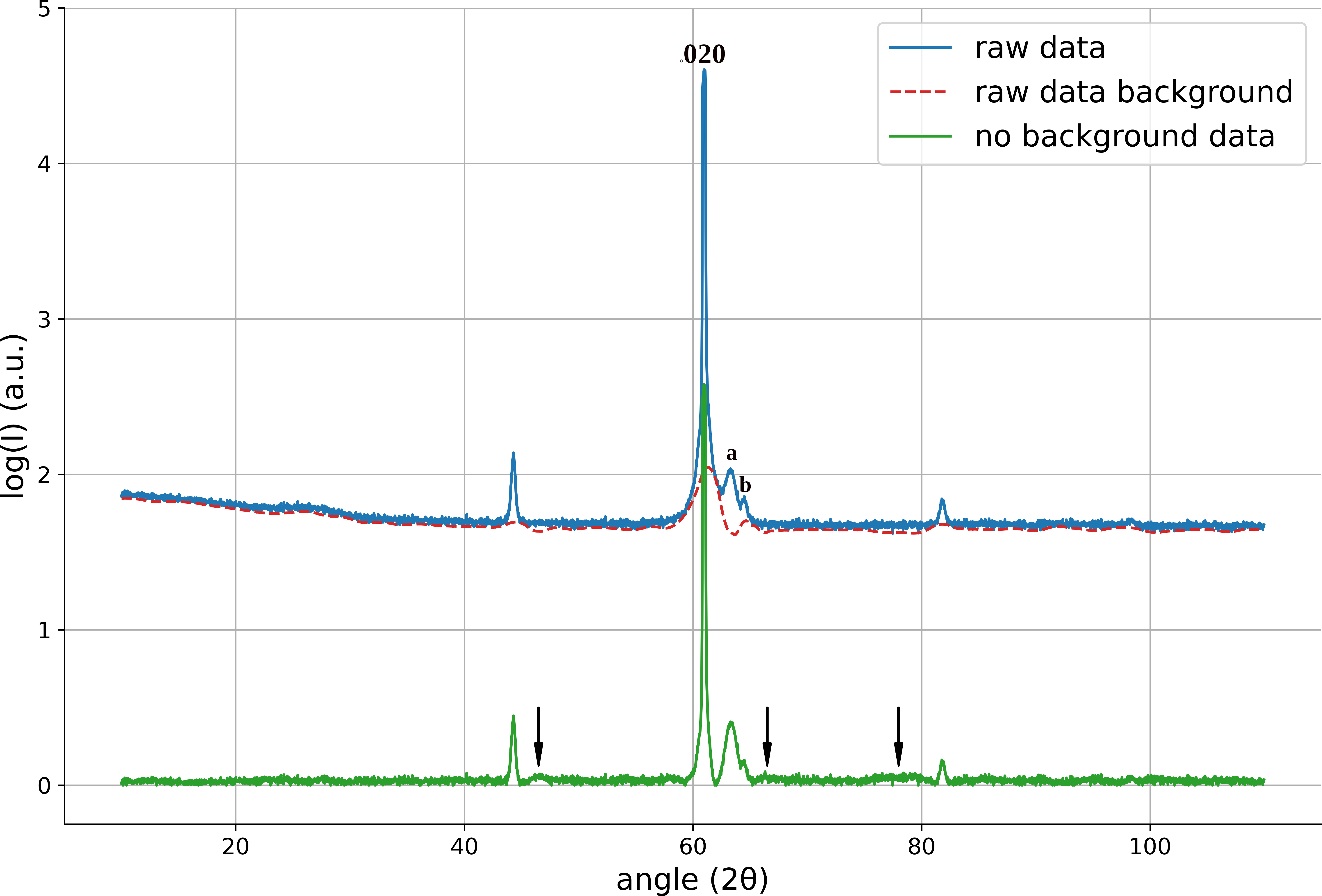}
	\caption{The logarithm of X-ray measured spectrum intensities from  (blue solid line) of $Ga_{2}O_{3}$ crystal, with well visible, high-level background; the spectrum with removed background is given as a green line. Arrows indicate artifacts remained after DWT synthesis (\textit{db5})}
	\label{fig:xrd-spectra}
\end{figure}

The background reduction and CWT analysis results of raw and processed data are shown for the X-ray powder diffraction spectrum of the $\beta-Ga_{2}O_{3}:Yb$ system in Fig. \ref{fig:xrd-spectra}. In the first step, the intensities of the raw spectrum have been reduced by calculating their logarithm $log_{10}$, to decrease the height ratio and enhance readability.  Next, the spectrum points have been processed by application of DTCWT algorithm based on \textit{db5} wavelet with decomposition up to the 6th level. 

The presented diffraction spectrum (represented by a solid blue line) exhibits a high degree of natural background, gradually diminishing towards the highest-valued angles. The calculated background, depicted as a dashed red line, accurately reproduces the natural course, exhibiting a high level of compliance and being free of medium and high-frequency components. It has been subtracted from the raw data, resulting in a final, free of slow-changing components spectrum (green line). The processed spectrum kept the general shape of the peaks.  In the case of the 020 peak, the algorithm detected and subsequently reduced the local excess of background, which also allowed the extraction of the initially weak peak (marked as \textbf{a}) around $63^{\circ}$. The neighbor peak \textbf{b} has remained untouched. The spectrum also consists of slightly lifted regions, indicated by the arrows, that have no physical meaning but are the remains of DWT synthesis only. 

\begin{figure}
	\centering
    \includegraphics[width=0.5\linewidth]{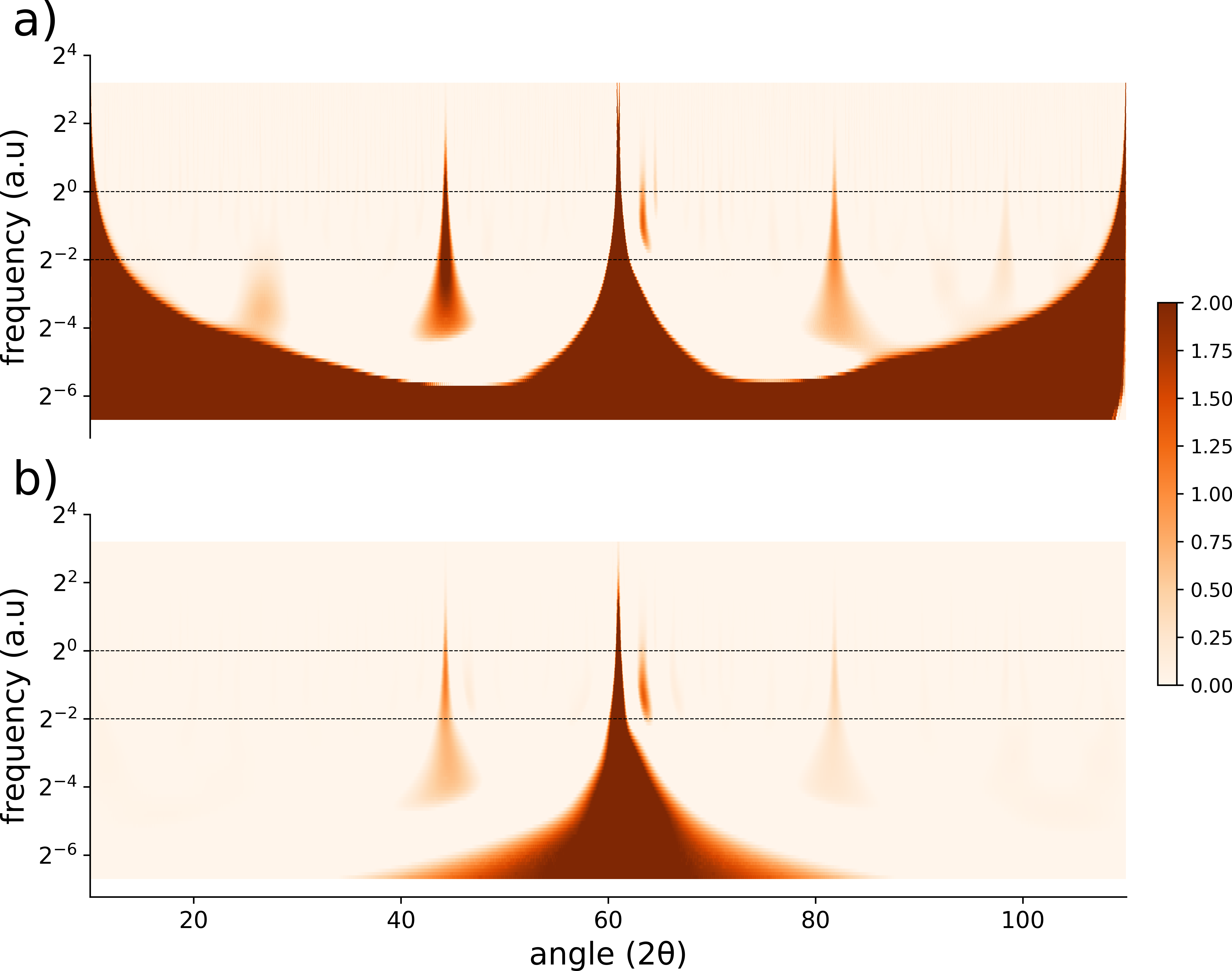}
	\caption{ CWT spectra of a) initial X-ray intensities and b) after slow changing components ($f<0.25$) reduction. To enhance readability, only CWT coefficients above 0 are shown; high-frequency ($f>1$) noise has been removed by means of SG filtering}
	\label{fig:xrd-cwt}
\end{figure}

For the above cases, CWT spectra have been calculated (see Fig.\ref{fig:xrd-cwt}) to show the impact of the DTCWT algorithm on the data. To enhance visibility, only positive CWT coefficients are shown, and high-frequency noise ($f>1\ a.u.$) has been removed; high-intensity regions are marked in dark-orange colors. The sampling frequency is equal to:

\begin{equation}
	f_{s}=\frac {number\ of \ data\ points}{angle\ range}\approx \frac{3808}{100} \approx 38.08
\end{equation}

The CWT spectrum of the initial XRD pattern (Fig.\ref{fig:xrd-cwt}a), shows that all angle ranges consist of a high number and significant intensity of low-frequencies ($f<0.25 a.u.$) components. The diffraction peaks are well localized on the angle axis and cover mainly the narrow range of medium frequencies, i.e., $0.25<f<1$. 

After background reduction (Fig.\ref{fig:xrd-cwt}b), the regions of low frequencies have been significantly reduced. However, in the case of the medium frequencies sub-band, the components corresponding to diffraction peaks are preserved and also partially preserve the region of very low frequencies, i.e., $f<2^{-6}=0.015625$ corresponding to the background under the 020 peak. One can also notice that the background reduction resulted in a decrease in overall intensities. 
In summary, the CWT clearly shows that the DTCWT algorithm efficiently recognizes and separates low frequencies from other frequency groups.

\begin{figure}
	\centering
    \includegraphics[width=0.9\linewidth]{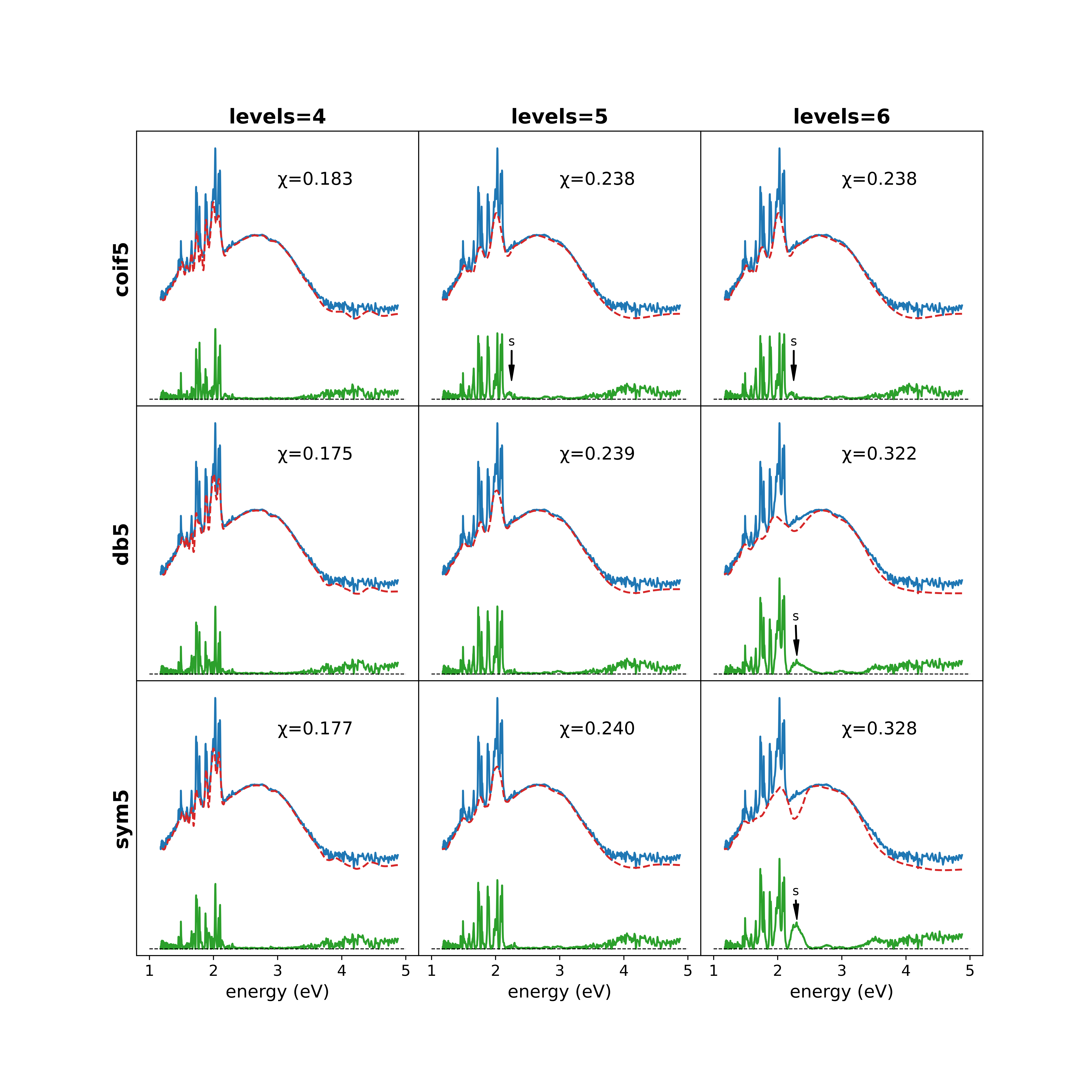}
	\caption{ Comparison of wavelet families and levels of decomposition for PL spectra. To enhance the visibility, the $log_{10}(I)$ of input data (blue line) has been processed; background and outcomes are given respectively as red and green lines}
	\label{fig:cmpDWT}
\end{figure}

As previously mentioned, the experimental data comprises numerous components with distinct characteristics. Therefore, input is processed in many and sometimes non-trivial ways to meet requirements. In the case of the DWT applications, one must only consider the selection of the wavelet family and the number of decomposition levels. The results of such analysis are demonstrated in Fig.\ref{fig:cmpDWT}, where the PL spectra of $Ga_2O_3:Eu$ system were used. To compare the effect of background reduction, the $\chi$-metric has been calculated as follows:
\begin{equation}
	\chi=\sqrt{\frac{ \sum_i^N (y_i-b_i)^2 } {N}}
\end{equation}
where: $y_i$ stands for input data; $b_i$ - background; $N$ - number of data points. The low values of $\chi$-metric demonstrate the background overfitting to input data and underfitting in the opposite.

Studies of PL spectra show a significant reduction in background, regardless of peak characteristics, overall background shape, and the presence of high-frequency noise. The common feature for all cases, independent of the selected family or the number of decomposition levels, is enhancing the high-frequency noise for energies above $3eV$. It is the most pronounced if decomposition is above the 4th level.

There is a visible dependence on wavelet family and decomposition level selection, which can interfere with peak magnitude and the presence of artifacts.

The selection of the wavelet family has a minor impact on the final results. There are only negligible differences between families seen for the highest level of decompositions. The essential factor for background reduction is the number of decomposition levels. Here, increasing the decomposition level clearly goes with increasing the value of $\chi$-metric. This dependence also indicates that for low $\chi$ values, i.e., about $0.18$, see in Fig.\ref{fig:cmpDWT} the first column, there is an unfavorable effect of background overfitting to data, especially for narrow peaks placed at energies near $2eV$, and gives a disruption in their height proportions. 

In contrast, the unfavorable underfitting effect is present at the highest level of decomposition, see in Fig.\ref{fig:cmpDWT} the last column, where the $\chi$ is about $0.32$. It induces the creation of spurious peaks \textit{s} (indicated by arrows) with significant height. For the \textit{db5} and \textit{sym5} families, one can notice a poorly fitted local background for real peaks at $E=2 eV$. The presence of \textit{s} illustrates the capability of discontinuity detection by wavelets, which is the source of error here.

An exception one can observe for \textit{coif5}, see in Fig.\ref{fig:cmpDWT}, the first row. In this case, both the 5th and 6th levels give exactly the same $\chi$, equal to $0.238$, well-separated peaks and spurious peaks with very low heights. This example demonstrates that decomposition based on Coiflet wavelets detects irrelevant differences for DTCWT without a significant negative impact on background reduction.

\section{Discussion and conclusions}

The above results demonstrate that DTCWT is a very effective tool for background removal from data with minimal impact on peak properties and user interventions. However, it requires more attention if the subject of analysis is a neighborhood of real peaks. In this case, wavelets generate spurious ripples around the main peaks, even if the selection of the family and number of decomposition levels is done correctly. The same effect is given if the number of decomposition levels is too high. In the case of coexistence of a wide, slowly changing background with a relatively short region of high-frequency noise, see Fig.\ref{fig:cmpDWT}, the presence of noise makes it difficult to find a correct level of background. This issue must be addressed by the high-frequency filtering, e.g., with the Savitzky-Golay method. 

It is shown that the most important for DTCWT is the proper selection of the number of decomposition levels. The number of levels should be close to the maximal value $L_{max}$ for a given set of data points. Here, the best trade-off has been found if the number of levels is equal to $L_{max}-1=5$.  

The choice of the wavelet family (WF) is of lesser importance. For X-ray diffraction, the most suitable family was \textit{db5}, while for PL spectra, both \textit{db5} and \textit{sym5} wavelet families. Slightly worse results were obtained for \textit{coif5} wavelets.

To conclude, the DTCWT algorithm requires fewer initial parameters for data processing than typical fitting methods. Moreover, it is resistant to high-frequency noise, enables the extraction of weak features, and separates regions of interest very well. The final results are reliable due to the small risk of numerical errors' introduction.

\medskip
The software for background removal, namely \textit{tlorem.py}, based on DTCWT, written in Python language, is available from\cite{tlorem}. It also provides CWT analysis supported by \textit{pywt} packages\cite{pywt} and data denoising based on \textit{SciPy} package \cite{2020SciPy-NMeth}. Other Python scripts and data used for the paper preparation are available upon request.

\

\

\textbf{Acknowledgements} \\

The research was co-funded by the NCN project UMO-2022/45/B/ST5/02810. The experimental work was supported by the Helmholtz-Zentrum Dresden-Rossendorf (20002208-ST, 21002661-ST, 21002663-ST)

\bibliographystyle{abbrv}
\bibliography{biblio}

\end{document}